\documentclass[12pt,english]{article}
\usepackage[latin9]{inputenc}
\usepackage{geometry}
\usepackage{color}
\usepackage{bm}
\usepackage{amstext}
\usepackage{setspace}
\usepackage{esint}
\usepackage[unicode=true,pdfusetitle,
 bookmarks=true,bookmarksnumbered=false,bookmarksopen=false,
 breaklinks=false,pdfborder={0 0 0},pdfborderstyle={},backref=false,colorlinks=true]
 {hyperref}
\hypersetup{
 linkcolor=cyan, citecolor=blue}
\usepackage{breakurl}

\makeatletter
\newcommand{\lyxaddress}[1]{
\par {\raggedright #1
\vspace{1.4em}
\noindent\par}
}

\makeatother

\begin{document}

\title{Thermal vorticity production in relativistic dissipative fluids \footnote{Presented by S. Pu at the Critical Point and Onset of Deconfinement Conference,
Wroclaw, Poland, May 30 - June 4, 2016.} \date{}}

\author{Yang-guang Yang$^{1}$ and Shi Pu$^{2}$}
\maketitle

\lyxaddress{\begin{center}
$^{1}$\emph{Department of Modern Physics, University of Science
and\\ Technology of China, Hefei, Anhui 230026, China}
\par\end{center}}

\lyxaddress{\begin{center}
$^{2}$\emph{Department of Physics, The University of Tokyo, \\7-3-1
Hongo, Bunkyo-ku, Tokyo 113-0033, Japan}
\par\end{center}}
\begin{abstract}
We have computed the circulation integrations of thermal vorticity
with and without charged currents in dissipative fluids. We find that
the relativistic Kelvin circulation theorem will be modified by the
dissipative effects, therefore, the circulation integrations of thermal
vorticity may not be conserved during the fluid evolution. 
\end{abstract}
Recently, some novel chiral transport phenomena in relativistic heavy
ion collisions have been extensively studied. In a Weyl fermionic
system with imbalanced chirality, charged currents can be induced
by magnetic or vortical fields, the so-called chiral magnetic or vortical
effect \cite{Kharzeev:2007jp,Fukushima:2008xe}, respectively. In
such a system, magnetic and vortical fields also enhance the separation
of chirality, which is called chiral separation effect \cite{Kharzeev:2007jp,Fukushima:2008xe}.
These phenomena can be described by chiral kinetic equations (CKE),
which are obtained by a variety of approaches, such as path integral
\cite{Stephanov:2012ki,Chen:2013iga,Chen:2014cla}, Hamiltonian \cite{Son:2012wh,Son:2012zy}
and quantum kinetic theory via Wigner functions \cite{Gao:2012ix,Chen:2012ca,Gao:2015zka,Hidaka:2016yjf,Gao:2017gfq},
see, for example, Ref. \cite{Bzdak:2012ia,Kharzeev:2013ffa,Kharzeev:2015kna}
for reviews. Similar effects in relativistic quantum system (2+1)-dimensional
can also be discussed by the quantum kinetic theory \cite{Chen:2013dca}.
In Refs. \cite{Huang:2013iia,Pu:2014cwa,Jiang:2014ura,Pu:2014fva},
in analogy to magnetic fields, the electric fields can also induce
a chiral current, the so-called chiral electric separation effect.
Similar to normal Hall effects, the chiral Hall effects and the induced
density waves are investigated in Ref. \cite{Pu:2014fva}. Some new
chiral transport effects from the electromagnetic fields up to the
second order in the gradient expansion in fluid dynamics have been
systemically studied in Refs. \cite{Chen:2016xtg,Gorbar:2016qfh,Gorbar:2016aov,Ebihara:2017suq}.

Since an extremely strong magnetic field of the order $\sim10^{18}-10^{19}G$
is produced in relativistic heavy ion collisions \cite{Kharzeev:2007jp,Deng:2012pc,Roy:2015coa,Li:2016tel},
magento-hydrodynamics with chiral transport effects may be an important
tool for these phenomena. Some progress has been made along this line.
The longitudinal boost-invariant Bjorken flow in (1+1)-dimension with
a transverse magnetic field with \cite{Pu:2016ayh} and without magnetization
effects \cite{Roy:2015kma}. Furthermore, a perturbative solution
for Bjorken flow in (2+1)-dimension with external magnetic fields
has been found in Ref. \cite{Pu:2016bxy}, (also see the recent simulations
Ref. \cite{Roy:2017yvg}).

The vorticial field can lead to the local polarization effect \cite{Gao:2012ix}.
The vorticity gives the polarization of hadrons at late time of heavy-ion
collisions \cite{Pang:2016igs,Fang:2016vpj}. Several groups estimate
the vorticity in different ways, e.g. by the AMPT \cite{Jiang:2016woz}
and HIJING model \cite{Deng:2016gyh}. In Refs. \cite{PhysRevLett.105.095005,Gao:2014coa},
the thermal vorticity generation in an ideal fluid is discussed. Without
charged particles, the circular integration of thermal vorticity is
proved to be conserved, which is called relativistic Kelvin circulation
theorem. In this note, we will follow Ref. \cite{Gao:2014coa} and
extend it to a dissipative fluid. We find that the dissipative effects
including viscosity and heat conductivity might induce thermal vorticity.

Without charged particles, the energy-momentum conservation reads,
\begin{equation}
\partial_{\mu}T^{\mu\nu}=0,\label{eq:consevation}
\end{equation}
where $T^{\mu\nu}$ is the energy-momentum tensor and is given by,
\begin{equation}
T^{\mu\nu}=(\varepsilon+P)u^{\mu}u^{\nu}-Pg^{u\nu}+\pi^{\mu\nu},
\end{equation}
where $\varepsilon$, $P$ are energy density and pressure, $u^{\mu}$
is the four-velocity of the fluid, and $\pi^{\mu\nu}$ is the viscous
tensor satisfying $u_{\mu}\pi^{\mu\nu}=0$. For simplicity, we only
discuss the fluid in the Landau frame, where the heat flux flow $h^{\mu}$
vanishes. Throughout this paper, we will use the convention for the
metric tensor $g_{\mu\nu}=g^{\mu\nu}=\textrm{diag}\{+,-,-,-\}$ and
the Levi-Civita tensor $\epsilon^{0123}=-\epsilon_{0123}=1$. Then,
we have $u^{\mu}$ satisfies $u^{2}=1$, and the orthogonal projector
is given by $\Delta^{\mu\nu}=g^{\mu\nu}-u^{\mu}u^{\nu}$. Using the
thermodynamic relations $\varepsilon+P=Ts$ and $d\varepsilon=Tds$,
with $s$ the entropy density and $T$ the temperature, and Eq. (\ref{eq:consevation})
in the local rest frame with $u^{\mu}=(1,\mathbf{0})$, a circulation
integral along a covariant loop $L(\tau)$ is given by, 
\begin{eqnarray}
\frac{d}{d\tau}\oint_{L(\tau)}Tu^{\mu}dx_{\mu} & = & -\oint_{L(\tau)}\frac{1}{s}\Delta^{\nu\beta}\partial^{\alpha}\pi_{\alpha\beta}dx_{\nu},\label{eq:flux_viscous_proper_01}
\end{eqnarray}
where $\tau$ is the proper time defined by $d/d\tau=u^{\mu}\partial_{\mu}$.
In the laboratory frame, the fluid velocity becomes $u^{\mu}=\gamma(1,\bm{v})$
with $\gamma=1/\sqrt{1-|\bm{v}|^{2}}$. A similar circulation integration
for the thermal vorticity along a loop $L(t)$ is, 
\begin{eqnarray}
 &  & \frac{d}{dt}\int_{S(t)}\left[\bm{\nabla}\times(T\gamma\bm{v})\right]\cdot d\bm{S}\nonumber \\
 & = & -\oint_{L(t)}\frac{1}{\gamma s}[-\gamma^{2}\bm{v}\partial^{\alpha}\pi_{\alpha0}+\gamma^{2}\bm{v}(v^{i}\partial^{\alpha}\pi_{\alpha i})]\cdot d\bm{x}+\oint_{L(t)}\frac{1}{\gamma s}\partial^{\alpha}\pi_{\alpha}^{i}dx_{i}.\label{eq:vorticity_lab_01}
\end{eqnarray}

Now let us consider a fluid with charged particles. In presence of
electromagnetic fields, the energy-momentum conservation and charge
conservation equations read, 
\begin{eqnarray}
\partial_{\mu}T^{\mu\nu} & = & F^{\nu\lambda}j_{\lambda},\ \partial_{\mu}j^{\mu}=0,\;\partial_{\mu}j_{i}^{\mu}=0,\label{eq:consevation-2}
\end{eqnarray}
where $j^{\mu}$ is the electric current, $j_{i}^{\mu}$ are other
conserved currents and $F^{\mu\nu}=\partial^{\mu}A^{\nu}-\partial^{\nu}A^{\mu}$
is the electromagnetic field strength. For example, if without chiral
anomaly, $j_{i}^{\mu}$ can be the chiral current in a Weyl fermionic
system. Generally, we can also write, 
\begin{eqnarray}
j_{i}^{\mu} & = & n_{i}u^{\mu}+\nu_{i}^{\mu},
\end{eqnarray}
where $n_{i}$ are the number densities for conserved charges and
$\nu_{i}^{\mu}$ are the heat currents. For the electric current,
the decomposition is similar, $j^{\mu}=nu^{\mu}+\nu^{\mu}$, with
$n$ the electric charge number density, and $\nu^{\mu}$ the heat
and electric conducting currents. The thermodynamic relation becomes
$\varepsilon+P=Ts+\mu n+\sum_{i}\mu_{i}n_{i}$, where $\mu$, $\mu_{i}$
are the chemical potentials. We follow Ref. \cite{Gao:2014coa} to
consider the enthalpy vorticity instead of thermal vorticity. We find
in the local rest frame,

\begin{eqnarray}
\frac{d}{d\tau}\oint_{L(\tau)}(fu^{\mu}+A^{\mu})dx_{\mu} & = & -\oint_{L(\tau)}\left\{ T\partial^{\mu}\left(\frac{s}{n}\right)+\sum_{i}\mu_{i}\partial^{\mu}\left(\frac{n_{i}}{n}\right)\right.\nonumber \\
 &  & \left.-\frac{1}{n}[fu^{\mu}(\partial\cdot\nu)+F^{\mu\lambda}\nu_{\lambda}-\partial_{\nu}\pi^{\nu\mu}]\right\} dx_{\mu},\label{eq:vorticity_LRF_02}
\end{eqnarray}
where $f=(\varepsilon+P)/n$ is the enthalpy density and in the laboratory
frame, 
\begin{eqnarray}
 &  & \frac{d}{dt}\int_{S(t)}\left[\bm{B}+\bm{\nabla}\times(f\gamma\bm{v})\right]\cdot d\bm{S}\nonumber \\
 & = & -\int_{S(t)}\left[\bm{\nabla}\left(\frac{T}{\gamma}\right)\times\bm{\nabla}\left(\frac{s}{n}\right)+\sum_{i}\bm{\nabla}\left(\text{\ensuremath{\frac{\mu_{i}}{\gamma}}}\right)\times\bm{\nabla}\left(\frac{n_{i}}{n}\right)\right.\nonumber \\
 &  & \left.+\bm{\nabla}\times\frac{1}{n}f\bm{v}(\partial\cdot\nu)\right]\cdot d\bm{S}-\oint_{L(t)}\frac{1}{\gamma n}\left[\partial_{\nu}\pi^{\nu i}-F^{i\lambda}\nu_{\lambda}\right]\cdot dx_{i}.\label{eq:vorticity-2-1}
\end{eqnarray}
where $\mathbf{B}$ is the magnetic field.

Equations (\ref{eq:flux_viscous_proper_01}, \ref{eq:vorticity_lab_01},
\ref{eq:vorticity_LRF_02}, \ref{eq:vorticity-2-1}) are our main
results for the circular integration of thermal vorticity in the local
rest and laboratory frame with and without conserved currents.

Now we will focus on a system in the absence of charged particles
in the local rest frame. In Eq. (\ref{eq:flux_viscous_proper_01}),
without dissipative effects, the circular integration $\oint_{L(\tau)}Tu^{\mu}dx_{\mu}$
is conserved, so-called the relativistic Kelvin circulation theorem
\cite{PhysRevLett.105.095005,Gao:2014coa}. It seems that the variation
of shear viscous tensor becomes a source to the circular integration
of the thermal vorticity. On the other hand, the entropy density in
this case is also not conserved. Therefore, we cannot conclude that
dissipative effects will violate the relativistic Kelvin circulation
theorem. For example, in a longitudinal boost-invariant Bjorken flow
in (1+1)-dimension with non-zero shear viscosity, the fluid is homogenous
in the transverse plane, i.e. both $T$ and $u^{\mu}$ are independent
on the space coordinates $(x,y)$ in the transverse direction. Therefore,
any circular integration $\oint_{L(\tau)}Tu^{\mu}dx_{\mu}$ along
a loop in the transverse plane will vanish during the evolution. However,
this result does not hold in a Bjorken flow in (3+1)-dimension. So
far, we conclude that circulation integration of thermal vorticity
may not be conserved due to the dissipative effects.

Then, let us briefly discuss a system in the presence of charged particles
in the local rest frame without shear viscous tensor. It is a little
surprising that the heat currents $\nu_{i}^{\mu}$ do not appear in
the right handed side of Eq. (\ref{eq:vorticity_LRF_02}), since the
electromagnetic fields only couple with electric current $j^{\mu}$.

To end this note, we summarize that we have obtained the dynamical
equations for the circular integration of thermal vorticity with dissipative
effects. In the further considerations, we will learn the generation
of thermal vorticity through other approaches, e.g. the kinetic theory
and hydrodynamic simulations.

\emph{Acknowledgment} The authors will thank Qun Wang and Jian-hua
Gao for the helpful discussions. S. P. was supported by JSPS post-doctoral
fellowship for foreign researchers.

\bibliographystyle{h-physrev} 
\bibliography{main}
\end{document}